First-principles calculation of the lattice thermal conductivities of α-, β-, and γ-Si$_3$N$_4$

Kazuyoshi Tatsumi[a,b,*], Atsushi Togo[b] and Isao Tanaka[b,c,d]

[a] *Advanced Measurement Technology Center, Institute of Materials and Systems for Sustainability, Nagoya University, Chikusa, Nagoya 464-8603, Japan*

[b] *Center for Elements Strategy Initiative for Structural Materials, Kyoto University, Sakyo, Kyoto 606-8501, Japan*

[c] *Department of Materials Science and Engineering, Kyoto University, Sakyo, Kyoto 606-8501, Japan*

[d] *Nanostructures Research Laboratory, Japan Fine Ceramics Center, Atsuta, Nagoya 456-8587, Japan*

[*] *To whom all correspondences should be addressed. E-mail: k-tatsumi@imass.nagoya-u.ac.jp*

Abstract

Lattice thermal conductivities (LTCs) of α-, β- and γ-Si$_3$N$_4$ single crystals are investigated from *ab initio* anharmonic lattice dynamics, within the single-mode relaxation-time approximation of the linearized phonon Boltzmann transport equation. At a temperature of 300 K, a $\kappa_{xx}$ of 70 and a $\kappa_{zz}$ of 98 (in units of Wm$^{-1}$K$^{-1}$) are obtained for α-Si$_3$N$_4$. For β-Si$_3$N$_4$, $\kappa_{xx}$ and $\kappa_{zz}$ are found to be 71 and 194, respectively, which are consistent with the reported experimental values of 69 and 180 for individual β-Si$_3$N$_4$ grains in a ceramic. The theoretical $\kappa_{xx}$ values of α- and β-Si$_3$N$_4$ are comparable, while the $\kappa_{zz}$ value of β-Si$_3$N$_4$ is almost twice that of α-Si$_3$N$_4$, which demonstrates the very large anisotropy in the LTC of the β phase. It is found that the large anisotropy in the LTC of β-Si$_3$N$_4$ was caused by the elongated Brillouin zone along the c$^*$ axis, where the acoustic phonons mostly contribute to LTC and have large group velocities even near the Brillouin zone boundary. The LTC of γ-Si$_3$N$_4$ is 81, which is as small as that of α-Si$_3$N$_4$, although γ-Si$_3$N$_4$ has much larger elastic constants. This means that elastic constants are not always a good indicator of LTC. We show that knowing the detailed distributions of both the group velocities and phonon lifetimes in the Brillouin zones is important for characterizing the LTC of the three phases.

I. INTRODUCTION

Several nitride insulators showing good thermal conductivities are important for heat sink materials used at elevated temperatures. Wurtzite-type w-AlN, which has an Adamantine (diamond-like) crystal structure, was noted by Slack et al. as exhibiting a large thermal conductivity of over 100 WK$^{-1}$m$^{-1}$.[1] Si$_3$N$_4$ has been more recently recognized as one of the good thermally conductive insulators. Remarkable advances in technologies related to the densification of the ceramic body and microstructural control have pushed the thermal conductivities of $Si_3N_4$ ceramics up to 177 $WK^{-1}m^{-1}$.[2-5] Since $Si_3N_4$ ceramics also exhibit high mechanical strength at elevated temperatures, they are regarded as ideal for use in various applications, such as engine components, gas turbines, and heat sink substrates of power semiconductor devices.

Under ordinary pressures, $Si_3N_4$ exists in one of two phases, α and β, both in the hexagonal lattice system, which are generally considered to be low- and high-temperature phases, respectively.[6,7] Their crystal structures are commonly formed by networks of $SiN_4$ tetrahedra, whose manners of stacking along the c axis are different. The unit cell periodicity of the α phase is approximately two times longer than that of the β phase, with lattice constants of c = 5.62[8] and 2.91[9] Å, respectively. In addition to the α and β phases, a high-pressure γ phase is known to form a cubic spinel-type crystal structure.[10]

By using high-resolution thermoreflectance microscopy, Li *et al.* reported the thermal conductivities of individual rod-shaped β-$Si_3N_4$ grains in a ceramic to be 69 and 180 $Wm^{-1}K^{-1}$ along the a and c axes, respectively, and thus revealed the large anisotropy in thermal conductivity.[11] Takahashi *et al.* recently developed a technique whereby β-$Si_3N_4$ grains are coated with graphene of relatively high magnetic susceptibility, enabling them to align their c axes along the external magnetic field.[12] Based on this large anisotropy in thermal conductivity, it was proposed that the textural structure of rod-shaped β-$Si_3N_4$ grains would increase their thermal conductivity to a level matching or exceeding that of w-AlN.

Although the fabrication of millimeter-sized β-$Si_3N_4$ single crystals has been reported[13], the thermal conductivity of no isolated single crystal of any $Si_3N_4$ phase has yet been experimentally determined. It was proposed that the anisotropy in the thermal conductivity of β-$Si_3N_4$ phase grains may not stem from the intrinsic crystal properties, but rather, from the selective removal of crystal defects along the c axis of the grains.[14] Theoretically, Hirosaki *et al.* estimated the room-temperature lattice thermal conductivities (LTCs) $\kappa_{xx}$ and $\kappa_{zz}$ of α-$Si_3N_4$ to be 105 and 225, and $\kappa_{xx}$ and $\kappa_{zz}$ of β-$Si_3N_4$ to be 170 and 450, respectively, by applying the Green-Kubo formulation to the molecular dynamics method with the interatomic potentials proposed by Vashishta *et al.*[15] The ratio of the LTCs along the a and c axes agreed well with the experimental results obtained by Li *et al.*; however, the absolute values were more than two times larger than the experimental results. While the thermal conductivity of polycrystalline $Si_3N_4$ has been significantly improved, our basic knowledge of, for example, the thermal conductivity tensors of the different crystal phases remains insufficient.

The purpose of the present study was to determine the intrinsic thermal conductivities of the three $Si_3N_4$ phases through *ab initio* LTC calculations[16], in particular focusing on the anisotropy in the LTC of the β phase. The reasons for this anisotropy were discussed from a microscopic viewpoint. Since the high-pressure phase (γ) shows a larger bulk modulus and higher hardness than the ambient-pressure α and β phases[17,18], we expected the γ phase to exhibit higher thermal conductivities. Thus, we investigated the γ phase in comparison with α and β.

II.  COMPUTATIONAL PROCEDURES

A. Phonon thermal conductivity calculation using third-order interatomic force constants

The theoretical LTCs were calculated with the phonon-phonon interaction calculation code PHONO3PY[19], while the harmonic phonon states were analyzed with the phonon calculation code PHONOPY[20]. The LTC tensors were calculated by solving the linearized Boltzmann transport equation (LBTE) within the single-mode relaxation time approximation (RTA). Letting $\lambda = (\mathbf{q}, p)$ specify the phonon mode of a phonon wave vector $\mathbf{q}$ and the phonon band index p, the relaxation time due to phonon-phonon scattering was obtained for each $\lambda$ as the inversion of the self-energy $\tau_{\lambda,\text{ph-ph}} = 1/2\Gamma_\lambda(\omega_\lambda)$. The self-energy is given by

$$\Gamma_\lambda(\omega_\lambda) = \frac{18}{\hbar^2} \sum_{\lambda'\lambda''} |\Phi_{-\lambda\lambda'\lambda''}|^2 \begin{Bmatrix} (n_{\lambda'} + n_{\lambda''} + 1)\delta(\omega_\lambda - \omega_{\lambda'} - \omega_{\lambda''}) + \\ (n_{\lambda'} - n_{\lambda''})[\delta(\omega_\lambda + \omega_{\lambda'} - \omega_{\lambda''}) - \delta(\omega_\lambda - \omega_{\lambda'} + \omega_{\lambda''})] \end{Bmatrix}, \quad (1)$$

where $\Phi_{-\lambda\lambda'\lambda''}$ was obtained from the third-order interatomic force constants (IFCs) calculated from the *ab initio* band calculation and the harmonic phonon states, and $n_\lambda$ is the Bose-Einstein distribution. The mode frequency $\omega_\lambda$ was taken from the harmonic phonon states solved with the second-order IFC of the *ab initio* band calculation. In order to compare the more realistic results of the theoretical LTC with the experimental data, the isotopic scattering effect due to the natural isotope distribution was taken into account according to second-order perturbation theory.[21] With the relaxation times of the phonon-phonon scattering and isotropic scattering, $\tau_{\lambda,\text{ph-ph}}$ and $\tau_{\lambda,\text{iso}}$, the total relaxation time for a phonon mode was assumed to be $\tau_\lambda^{-1} \approx \tau_{\lambda,\text{ph-ph}}^{-1} + \tau_{\lambda,\text{iso}}^{-1}$.

Finally, the LTC tensors were obtained via

$$\boldsymbol{\kappa}(T) = \frac{1}{N_q \Omega} \sum_\lambda \tau_\lambda(T) \mathbf{v}_\lambda \otimes \mathbf{v}_\lambda c_\lambda(T), \quad (2)$$

where $T$ is the temperature, $\mathbf{v}_\lambda = \nabla_\mathbf{q} \omega_\lambda$ is the group velocity for the harmonic phonon mode, $N_q$ is the number of $\mathbf{q}$, $\Omega$ is the unit cell volume, and $c_\lambda$ is the mode heat capacity. To analyze the theoretical LTC in detail, we define the cumulative sum of the mode LTC below an $\omega$ of interest, $\boldsymbol{\kappa}^c(\omega)$:

$$\kappa^c(\omega) = \frac{1}{N_q \Omega} \int_0^\omega \sum_\lambda \tau_\lambda(T) \mathbf{v}_\lambda \otimes \mathbf{v}_\lambda c_\lambda(T) \delta(\omega' - \omega_\lambda) d\omega'. \qquad (3)$$

B.  Detailed conditions of *ab initio* band calculation

The IFCs were calculated using the first-principles projector augmented wave method (VASP code)[22]. The generalized gradient approximation of Perdew, Burke, and Ernzerhof[23] was used for the exchange correlation potential. A plane wave energy cutoff of 500 eV was employed. The crystal structures of the stationary structures were optimized until the convergence in the residual forces acting on the constituent atoms were less than $10^{-6}$ eV/Å. Supercell and finite difference approaches were used to calculate the IFCs.[24] The 1×1×2, 1×1×3, and 1×1×1 supercells of the conventional unit cells were adopted for the third-order IFCs of the α, β, and γ phases, respectively, while the larger supercells 2×2×2, 2×2×4, and 2×2×2 were adopted for the second-order IFCs; in both cases, the displacement amplitude was set to 0.03 Å. Table 1 shows the theoretical LTC values for several different sizes of supercells, indicating that the convergence of the theoretical LTC with respect to the supercell size is 2%.

Uniform **k**−point sampling meshes of 4×4×2, 4×4×3, and 3×3×3 were used for the third-order IFCs of the α, β, and γ phases, where the center of the $a^*$-$b^*$ plane (asterisks denote the reciprocal space) was shifted by 0.5 $a^*$ and 0.5 $b^*$ for α and β, respectively. For the second-order IFC, the Γ−point was only sampled for the α and β phase supercells and the (0.5, 0.5, 0.5) for the γ phase supercell. **q**−point sampling meshes of 14×14×16, 14×14×32, and 22×22×22 were used to calculate the LTCs in Eq. (2) for the α, β, and γ phases.

III.    LATTICE THERMAL CONDUCTIVITY CALCULATION RESULTS

In Table 2, the theoretical LTCs at 300 K are compared with the experimentally reported values, where it can be seen that the theoretical LTCs of β-$Si_3N_4$ are markedly more anisotropic than those of the α phase. The theoretical LTCs of β-$Si_3N_4$ are in good agreement with the corresponding experimental data for individual grains reported by Li *et al.*, which indicates that the experimentally reported large anisotropy in the thermal conductivities of β-$Si_3N_4$ stems from the intrinsic properties of the crystal, rather than specific defects induced during the sample preparation process. Among the nitrides studied, the theoretical LTC of β-$Si_3N_4$ along the c axis (194 $WK^{-1}m^{-1}$) is the closest to the values for high-thermal-conductivity AlN. Despite the high bulk modulus of the γ phase (Table 2), its theoretical LTC was as low as that of the α phase.

Fig. 1 shows the theoretical LTCs of the α and β phases as a function of *T*, in comparison with the reference experimental data. The temperature dependence of LTC can be approximated by $T^1$, because for these phases, a temperature of 300 K is high enough to approximate lifetimes by a high-temperature limit. The experimental data can be fitted reasonably well with curves (dashed curves in the figure) of the form $aT^{-1}+b$, where $a$ and $b$ are parameters. The experimental values of the β phase ceramic bulk[5] fall well between the calculated values of $\kappa_{xx}$ and $\kappa_{zz}$. If we compare the experimental values with $\sum_i \frac{\kappa_{ii}}{3}$, which is a simple directional average, the calculation shows slight underestimations with respect to the experiment, which can be understood from an experimentally tailored microstructure containing large β-$Si_3N_4$ grains selectively grown along the c axis[5].

For the α phase, the theoretical LTCs are reasonably close to the experimental values for a polycrystalline film, although the theoretical LTC curve decreases more steeply with increasing temperature. The thermal resistance is dominated by phonon-phonon interactions at high temperatures, but the thin-film synthesis process might induce possible lattice defects causing temperature-insensitive thermal resistance in the film.

Fig. 2 shows the $\kappa^c(\omega)$ of the β phase along two inequivalent directions and their ratios as an indicator of the anisotropic thermal conductivities; isotope effects are not considered in this figure. The cumulative sums of mode LTC $\kappa^c(\omega)$ indicate the frequency range over which the phonon modes contribute significantly to LTC. We can clearly see that the phonon modes in the frequency range of 0–12 THz contribute significantly to LTC. Interestingly, in the same frequency range, the anisotropy of $\kappa^c(\omega)$ increases monotonically. Thus, we can focus on this frequency range when we investigate the anisotropy of the thermal conductivity of the β phase in detail.

IV. DISCUSSION

A. Distributions of phonon frequencies in Brillouin zones of α and β phases

The cross-sections of the phonon frequency distributions on the $b^*c^*$ planes in the first Brillouin zones are shown in Fig. 3. The cross-sections of other planes containing the $c^*$ axis did not differ much from Fig. 3; thus, we focus on the results for the $b^*c^*$ planes as representative of all such planes. We show only the frequencies of four modes from the lowest frequency because they contribute significantly to LTC. In Fig. 3-b of the β phase, the iso-frequency lines in 0.06 Å$^{-1}$ $\leq q_{c^*} \leq 0.12$ Å$^{-1}$ are almost parallel to the $q_{b^*}$ axis, i.e., the group velocities in this section are nearly along the c axis. The cross-sections in Fig. 3-b indicate that the four modes of the β phase, in a significantly large part of the phonon states in the Brillouin zone, have group velocities oriented along the c axis, resulting in a much larger LTC along the c axis than along the a(b) axis in this phase. Fig. 3-a shows that the frequency distributions and group velocities of α-$Si_3N_4$ are fairly isotropic.

B. Harmonic phonon states of α, β and γ phases

Figure 4 shows the phonon band diagrams and density of states (DOS) of the three $Si_3N_4$ phases. They are almost identical to those reported ealier[27]. In the band diagram of β (Fig. 4-b), along the Γ-A path, the acoustic phonon branches highlighted in red increase their frequencies almost linearly from the Γ-point to the zone boundary of the A-point around 10 THz. Their gradients are large, so the group velocity components along the c axis maintain high values. Normally, optical branches are flat; however, the upper branches along the same path, highlighted in blue, also have significantly large group velocity components along the c axis.

In contrast, the corresponding acoustic branches in the α phase highlighted in red in Fig. 4-a do not increase their frequencies as much as those of the β phase. This is because the Γ-A path length of the α phase is approximately half that of β. The lattice constant c of α is nearly twice that of β, owing to the difference in the stacking of the $SiN_4$ tetrahedra along the c axis. The frequency maxima along the Γ-A path are around 7 THz, rather close to the maxima along the Γ-K and Γ-M paths (around 5 THz). Moreover, the upper branches highlighted in blue are relatively flat, like the upper branches along the Γ-K and Γ-M paths.

The total DOS of the α phase shows a distinct peak at ~6 THz, as indicated by an arrow, where the phonon branches become dispersionless over the entire path in the band diagram. The first peaks of α and β are related to the L-points or the Brillouin zone (BZ) edges with $c^* = 0.5$. The difference in their frequencies stems from difference in length between their BZs along $c^*$.

In the DOS projected onto the c and a axes (PDOS) shown in the insets of the total DOS, the two PDOS spectra of the β phase have a larger difference than in the case of α, which is consistent with the differences in the dispersion curves.

In the band diagram of the γ phase (Fig. 4-c), the acoustic phonon branches highlighted in red show significant linear dispersion along the L-Γ-X path. The frequencies of the longitudinal acoustic modes are 14 and 12.5 THz at the L- and X-points. The frequencies of the transverse acoustic modes are approximately half the values of the longitudinal modes at the L-point and a factor of $1/\sqrt{2}$ smaller than the longitudinal modes at the X-point, as in the case of fcc rare gas solids[27]. The roughly constant gradients of the branches are large, reflecting the large elastic constants of the γ phase. The number of phonon states in the frequency range of 0–5 THz, where only acoustic modes exist, is thus rather small since the DOS is small, which is why the γ phase has a lower thermal conductivity than β. By fitting the total DOS with a Debye model function, $a_D\omega^2$, the coefficient $a_D$ is found to be $11\times10^{-3}$, $8.6\times10^{-3}$, and $3.9\times10^{-3}$ ($THz^{-3} formula^{-1}$) for the α, β, and γ phases, respectively.

C.  Distributions of phonon lifetimes and their impact on the LTCs of the three phases

Although the anisotropy in LTC is simply explained by the harmonic phonon states, specifically, the group velocities, the absolute values of LTC among the three phases can be affected by phonon lifetimes, which is the focus of this section.

Fig. 5 compares phonon lifetimes as a function of ω. The DOS of the three phases are attached in the figure to allow the distributions of the phonon states to be examined with ease. The common trend is that the phonon lifetime decreases with increasing ω in the lower frequency region (< 6 THz). Most of the phonon lifetimes of the α phase are larger than those of the β phase in this frequency region. However, above 6 THz, the phonon lifetimes of the three phases are distributed similarly and roughly constant. The values for γ are as low as half those for α and β.

These differences in the distributions of the phonon lifetimes on the frequency axis affect the cumulative sums $\mathbf{\kappa}^c(\omega)$. Fig. 6-a plots $\mathbf{\kappa}^c(\omega)$ in the frequency range of 0–12 THz for the three phases. To see the importance of $\tau_\lambda$ in LTC, we approximate $\mathbf{\kappa}^c(\omega)$ using a constant phonon lifetime τ = 14 ps and a constant mode heat capacity $k_B$:

$$\mathbf{\kappa}^{c\prime}(\omega) = \frac{k_B \tau}{N_q \Omega} \int_0^\omega \sum_\lambda \mathbf{v}_\lambda \otimes \mathbf{v}_\lambda \delta(\omega' - \omega_\lambda) d\omega'. \tag{4}$$

This model could potentially reproduce the tendencies of the $\mathbf{\kappa}^c(\omega)$ curves; however, it fails owing to the different impact of phonon lifetimes in the three phases. These curves for the three phases are plotted in Fig. 5-b.

The mode heat capacity was set to Boltzmann's constant, $k_B$, as the high-temperature approximation, which is a reasonable approximation for ω < 19 THz, and we set τ to 14 ps and let the $\mathbf{\kappa}^{c\prime}(\omega)$ of the β phase approach $\mathbf{\kappa}^c(\omega)$ at ω = 12 THz for simplicity. Comparing $\mathbf{\kappa}^c(\omega)$ and $\mathbf{\kappa}^{c\prime}(\omega)$ on the high-frequency side, $\mathbf{\kappa}^{c\prime}(\omega)$ overestimates (underestimates) $\mathbf{\kappa}^c(\omega)$ in the γ (α) phase, because of the shorter (longer) $\tau_\lambda$ of the γ (α) phase above (below) 6 THz, indicating that prediction of the absolute LTC values in the present system requires a more accurate description of the anharmonicities. The γ phase shows a relatively small LTC because of the shorter $\tau_\lambda$ at ω > 6 THz, where the γ phase has relatively large numbers of the acoustic phonon states owing to the small $a_D$. The number of the acoustic phonon states above 6 THz can be estimated from $\int_{6\text{THz}}^{v_D} a_D \omega^2 d\omega$, where $v_D = \left(\frac{9}{a_D}\right)^{1/3}$ are the maximum frequencies of the Debye model. The numbers of acoustic phonon states are 0, 1.7, and 2.4 for the α, β, and γ phases, respectively.

The difference in $\mathbf{\kappa}^c(\omega)$ between the two directions of the β phase is well represented by $\mathbf{\kappa}^{c\prime}(\omega)$, confirming that the anisotropic LTC can be well explained by the anisotropic group velocities of the harmonic phonon states in the β phase. On the other hand, the gradients of both the

$\kappa^c(\omega)$ and $\kappa^{c'}(\omega)$ curves for the α phase decrease around ω = 6 THz, where the acoustic modes reach the Brillouin zone boundaries in all directions, resulting in less anisotropic LTC values.

V. SUMMARY

In the present study, we investigated the theoretical LTCs of the three $Si_3N_4$ phases by using phonon and phonon-phonon interaction calculations based on *ab initio* IFCs. The results can be summarized as follows:

1) The theoretical LTC of the β phase along the c axis was approximately three times as large as that along the a axis, in accordance with the experimental orientation dependence of the thermal conductivities of the individual grains, confirming that the anisotropic thermal conductivity of the β phase was due to the intrinsic crystal structure, rather than specific lattice imperfections of thermal resistance.
2) In contrast, the theoretical LTC of the α phase did not show significant anisotropy. The values averaged over all directions were larger for the β phase than for the α phase, which roughly agreed with the thermal conductivities reported for polycrystalline materials.
3) The significantly large LTC along the c axis in the β phase was explained by the theoretical observation that group velocities having large components along the c axis were more dominant in the acoustic phonon modes, compared to the α phase. The anisotropic group velocities in the β phase were due to the elongated Brillouin zone along the c* axis, with a shorter lattice constant c.
4) Although the group velocities of the acoustic mode of the γ phase was large, which was consistent with its large elastic constants, the LTC value was as low as that of the α phase because of the relatively large number of acoustic phonon states in the high-frequency region where the phonon lifetimes were relatively small.

ACKNOWLEDGEMENTS
The present work was partly supported by Grants-in-Aid for Scientific Research of MEXT, Japan (Grant No. 15K14108 and ESISM (Elements Strategy Initiative for Structural Materials) of Kyoto University).

FIGURE CAPTIONS

Fig. 1 Theoretical LTC curves of α- and β-Si$_3$N$_4$, compared with the experimental data. The dashed curves are the least-squares fits of the experimental data to $aT^{-1}+b$.

Fig. 2 Theoretical cumulative sums of mode LTC $\boldsymbol{\kappa}^c(\omega)$ (see Eq. 3) of β-Si$_3$N$_4$ along the a and c axes, i.e., $\kappa^c_{xx}(\omega)$ and $\kappa^c_{zz}(\omega)$, respectively, and their ratios, $\kappa^c_{zz}(\omega)/\kappa^c_{xx}(\omega)$.

Fig. 3 Contour maps of phonon frequency (THz) on the b$^*$c$^*$ planes of Brillouin zones. The maps for the four lowest-frequency phonon states are shown. The frequency landscapes are formed by simply connecting the frequencies of the same band indices, assigned by ascending order of frequency at the respective **q** points.

Fig. 4 Brillouin Zones (left), calculated phonon band diagrams (middle) and density of states (DOS) (right) for the three Si$_3$N$_4$ phases. The insets in DOS show DOS projected onto the a and c axes.

Fig. 5 Theoretical phonon lifetimes (top) and DOS (bottom) of the three phases.

Fig. 6 (a) Cumulative sums of mode LTC for the three phases $\boldsymbol{\kappa}^c(\omega)$ and (b) hypothetical cumulative sums with a constant phonon lifetime and a constant mode heat capacity $\boldsymbol{\kappa}^{c\prime}(\omega)$.

Table 1. Calculated lattice thermal conductivities of α, β, and γ-Si$_3$N$_4$ at 300 K with respect to several combinations of supercell sizes.

| Phase | Supercell (# of atoms) | | LTC (WK$^{-1}$m$^{-1}$) | |
|---|---|---|---|---|
| | 3$^{rd}$ FC | 2$^{nd}$ FC | xx | zz |
| α | 1x1x1 (28) | 1x1x1 (28) | 37 | 57 |
| | 1x1x2 (56) | 1x1x2 (56) | 41 | 79 |
| | 1x1x1 (28) | 2x2x2 (224) | 56 | 81 |
| | 1x1x2 (56) | 2x2x2 (224) | 70 | 98 |
| | 1x1x2 (56) | 2x2x3 (336) | 69 | 97 |
| β | 1x1x2 (28) | 1x1x2 (28) | 40 | 166 |
| | 1x1x2 (28) | 2x2x4 (224) | 75 | 208 |
| | 1x1x3 (42) | 2x2x4 (224) | 71 | 194 |
| | 1x1x3 (42) | 2x2x5 (280) | 72 | 197 |
| γ | 1x1x1 (56) | 1x1x1 (56) | 75 | |
| | 1x1x1 (56) | 2x2x2 (448) | 81 | |
| | 1x1x1 (56) | 3x3x3 (1512) | 82 | |

Table 2. Calculated thermal conductivities at 300 K, compared with the experimental data. Theoretical bulk moduli $B$ in units of GPa, calculated by the present band method, are presented in the fourth column.

|  | This work ||| Ref. Expt. |
| --- | --- | --- | --- | --- |
|  | $\kappa_{xx}$ | $\kappa_{zz}$ | $B$ | $\kappa$ |
| α-Si$_3$N$_4$ (trigonal) | 70 | 98 | 224 | 59[a] |
| β-Si$_3$N$_4$ (trigonal) | 71 | 194 | 237 | $\kappa_{xx}$:69[b]  $\kappa_{zz}$:180[b] |
| γ-Si$_3$N$_4$ (cubic) | 81 |  | 296 | --- |
| w-AlN (hexagonal) | 218 | 190 | 196 | 285[c] |

[a]Reference 25, poly-crystals.
[b]Reference 11, single crystalline grains of poly-crystals
[c]Reference 26, poly-crystals.

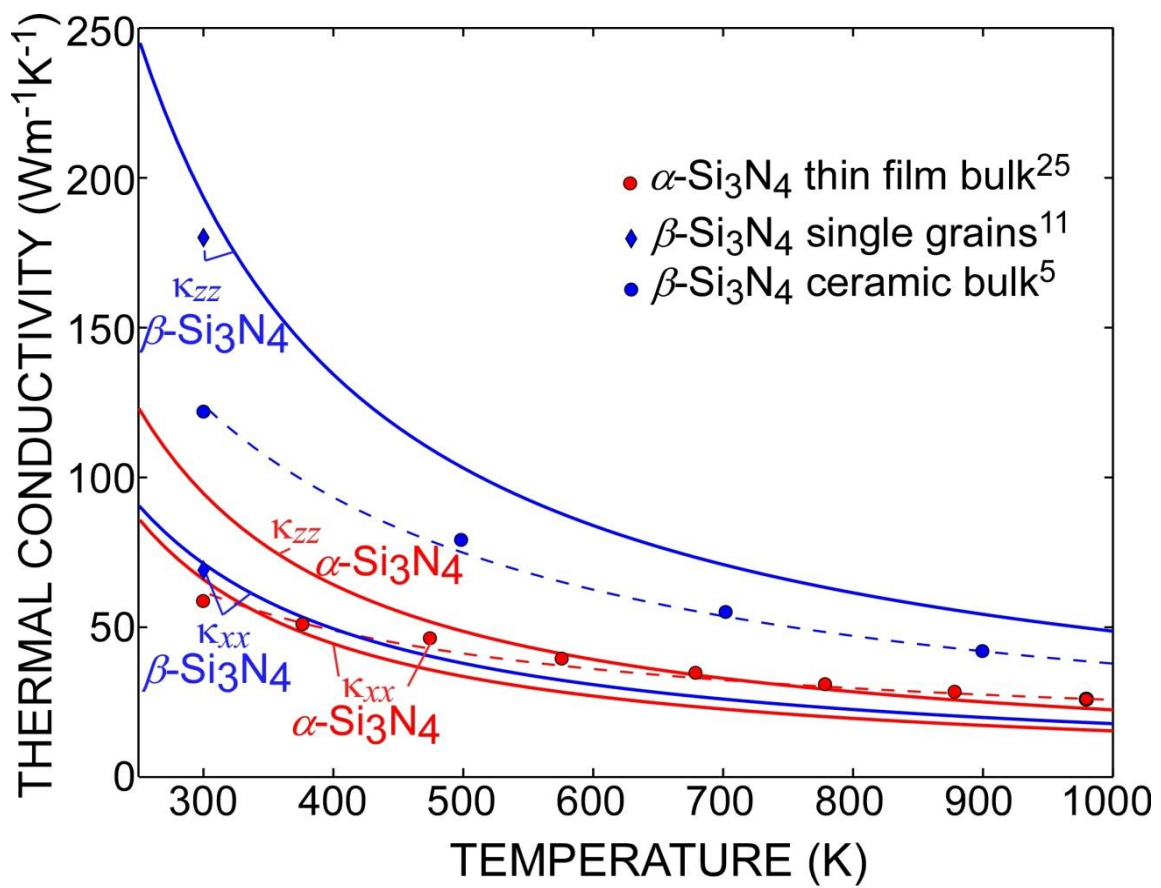

Fig. 1 Theoretical LTC curves of α- and β-Si$_3$N$_4$, compared with the experimental data. The dashed curves are the least square fits of the experimental data to $aT^{-1}+b$.

K. Tatsumi *et al.*

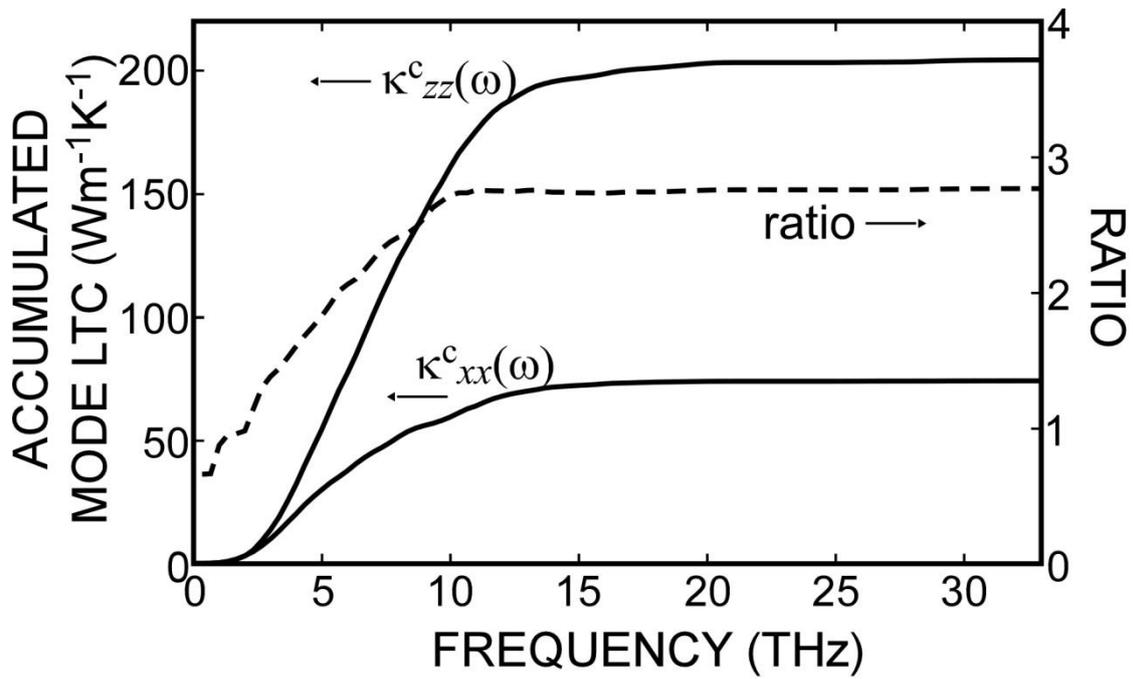

Fig.2 Theoretical cumulative sums of mode LTC $\kappa^c(\omega)$ (see Eq. 3) of $\beta$-$Si_3N_4$ along the a and c axes, i.e., $\kappa^c_{xx}(\omega)$ and $\kappa^c_{zz}(\omega)$, respectively, and their ratios, $\kappa^c_{zz}(\omega)/\kappa^c_{xx}(\omega)$.

K. Tatsumi *et al.*

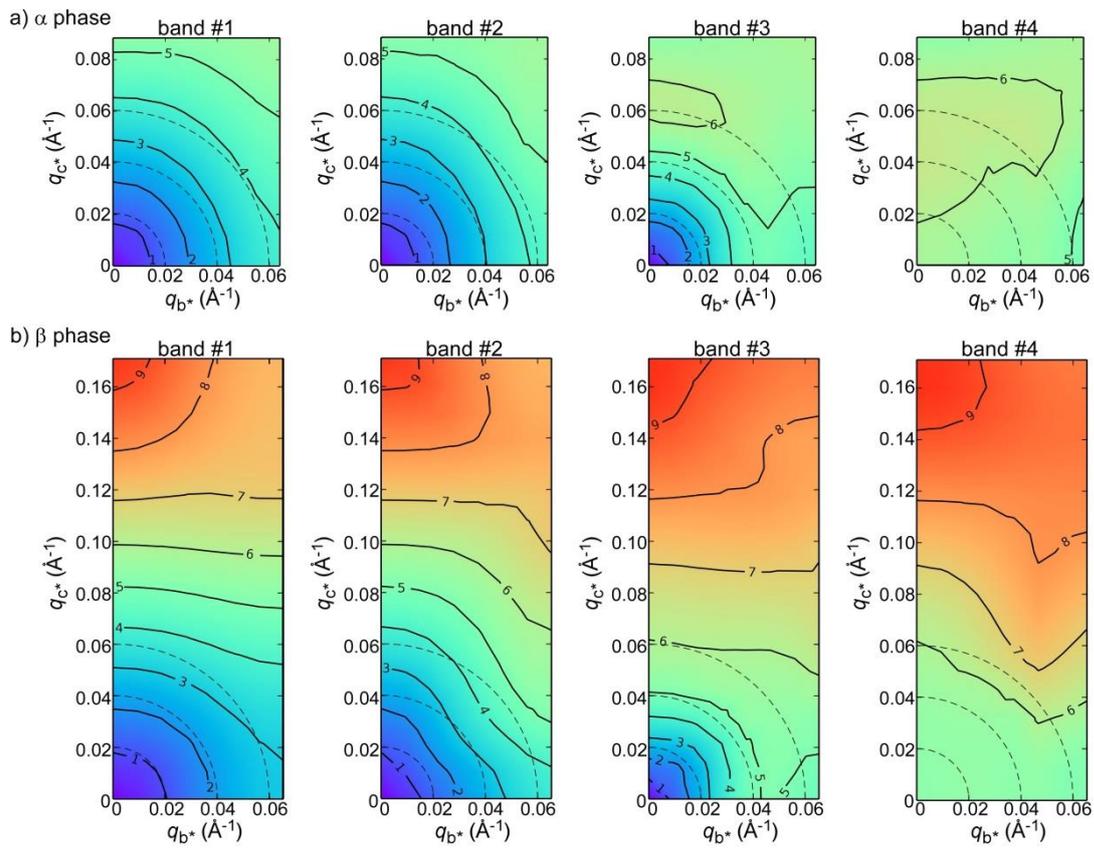

Fig. 3 Contour maps of phonon frequency (THz) on the b*c* planes of Brillouin zones. The maps for the four lowest-frequency phonon states are shown. The frequency landscapes are formed by simply connecting the frequencies of the same band indices, assigned by ascending order of frequency at the respective **q** points.

K. Tatsumi *et al.*

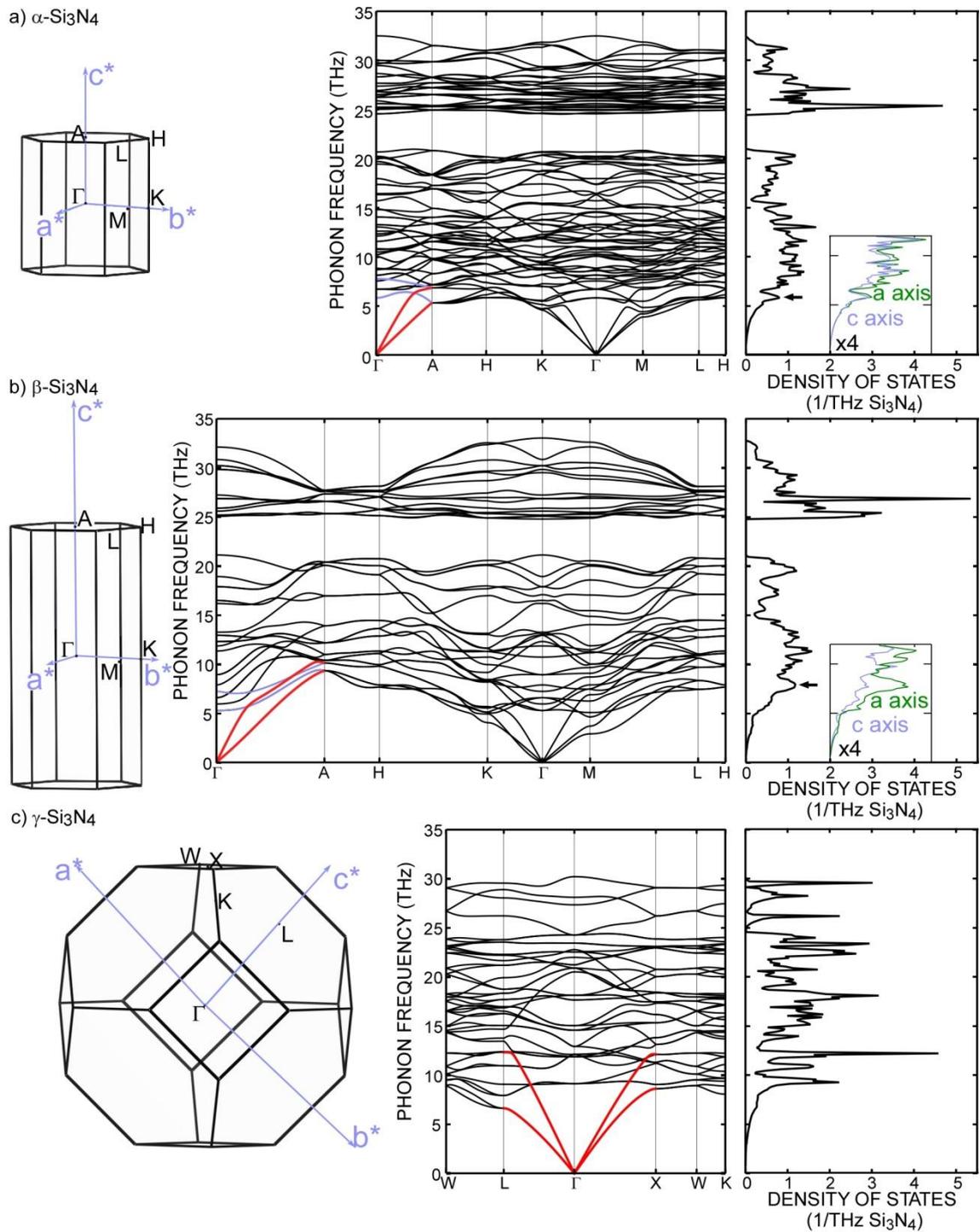

Fig.4 Brillouin Zones (left), calculated phonon band diagrams (middle) and density of states (DOS) (right) for the three Si$_3$N$_4$ phases. The insets in DOS show DOS projected onto the a and c axes.

K. Tatsumi *et al.*

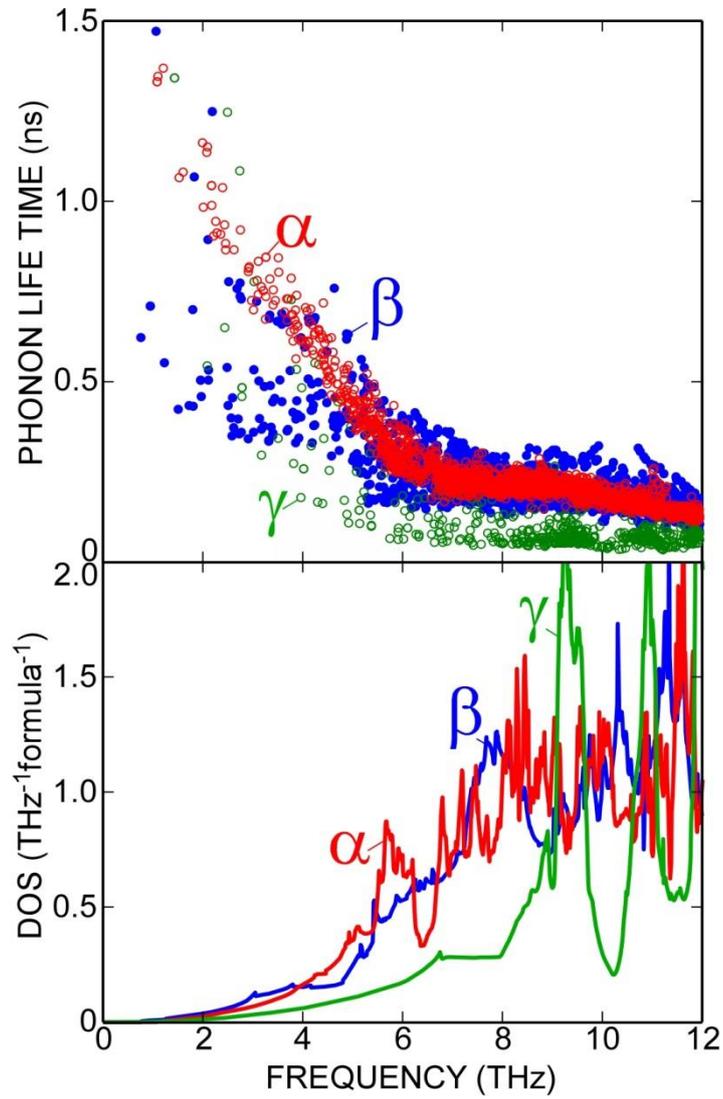

Fig.5 Theoretical phonon lifetimes (top) and DOS (bottom) of the three phases.

K. Tatsumi *et al.*

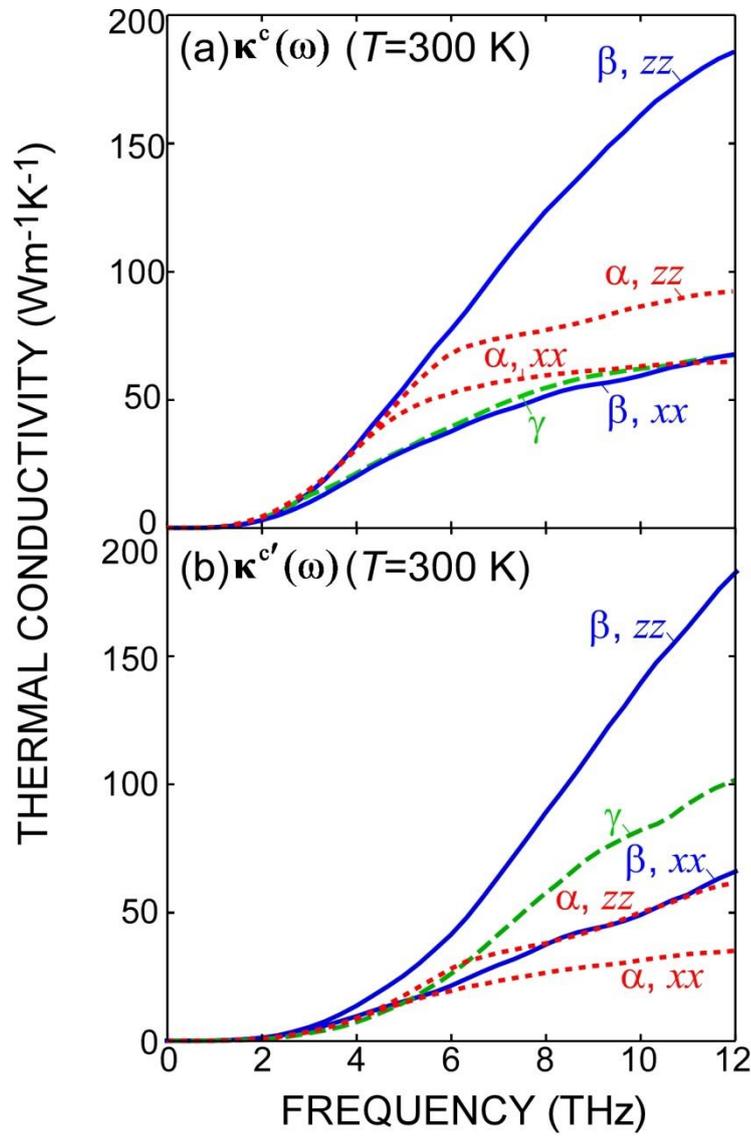

Fig.6 Cumulative sums of mode LTC for the three phases $\kappa^c(\omega)$ and (b) hypothetical cumulative sums with a constant phonon lifetime and a constant mode heat capacity $\kappa^{c'}(\omega)$.

K. Tatsumi *et al.*